\documentclass[aps,prd,10pt,preprintnumbers,superscriptaddress,showpacs,nofootinbib]{revtex4}
\usepackage{epsfig}
\usepackage{psfrag}
\usepackage{amsfonts}
\usepackage{graphicx}
\usepackage{dcolumn}
\usepackage{bm}
\usepackage{amsmath}
\usepackage{relsize}

\begin{document}

\title{Holographic spin alignment of vector mesons in a spinning black-hole background}

\author{Fei Wang}
\affiliation{School of Mathematics and Physics, China University
of Geosciences (Wuhan), Wuhan 430074, China}

\author{Zi-qiang Zhang}
\email{zhangzq@cug.edu.cn} \affiliation{School of Mathematics and
Physics, China University of Geosciences (Wuhan), Wuhan 430074,
China}

\begin{abstract}
Holographic predictions for vector-meson spin alignment may depend
on the background geometry and on the meson model. We examine the
opposite polarization preferences observed for charmonium and
light vector mesons by placing \(J/\psi\) and \(\phi\) in one
soft-wall model on the planar limit of a five-dimensional
equal-spin Myers--Perry AdS black hole. Spectral functions and
spin alignment are computed within the same current-correlator
framework; the nonrotating results of Sheng et al. [Phys. Rev. D
110, 056047 (2024)] are recovered at \(a=0\). At
\(T=0.15~\mathrm{GeV}\), \(J/\psi\) retains a clear resonance
peak, whereas \(\phi\) shows no narrow peak and instead behaves as
a broad distribution consistent with unstable \(s\bar{s}\) pairs.
For \(J/\psi\) along \(z\), rotation enhances the positive
deviation of \(\rho_{00}^{h}\) from \(1/3\) at low and
intermediate momenta but suppresses it at high momenta; along
\(x\), the effect is small. For \(\phi\), rotation enhances the
negative deviation, more along \(z\) than along \(x\). The
opposite helicity-frame signs and the opposing azimuthal and
rapidity trends of the global alignment provide a reference for
interpreting data: trends inconsistent with these would challenge
the simplest single homogeneous rotating-medium interpretation. We
distinguish model-dependent numerical details from qualitative
features stable within this model class.
\end{abstract}


\maketitle

\section{Introduction}
\label{sec:introduction}

Relativistic heavy-ion collisions produce a deconfined phase of
matter at high temperature and energy density, the quark-gluon
plasma
(QGP)~\cite{Bass_JPG1999,Scherer_PPNP1999,Shuryak_PPNP2009,Adams_NPA2005,Adcox_NPA2005,Shuryak_NPA2005}.
Heavy quarkonia, such as charmonium (\(c\bar c\)) and bottomonium
(\(b\bar b\)), are among the standard probes of this
medium~\cite{CasalderreySolana_arXiv2011}. Because the medium is
strongly coupled, nonperturbative methods are essential. The
anti-de Sitter/conformal field theory (AdS/CFT)
correspondence~\cite{Maldacena_ATMP1998,Gubser_PLB1998,Witten_ATMP1998,Aharony_PhysRep2000}
has provided such a method for strongly coupled plasmas, and
related holographic studies have examined heavy-quarkonium spectral
functions and thermal
dissociation~\cite{Zhao_EPJC2022,Zhao_EPJC2023,Braga_PLB2017,Braga_PLB2018,Mamani_PRD2022,Braga_PRD2023,Zhu_CPC2025,Braga_EPJC2016,MartinContreras_PRD2021,Toniato_PRD2025,Zhu_PRD2024},
thermal
widths~\cite{Zhao_EPJA2020,Feng_PRD2020,Braga_PRD2016,Finazzo_JHEP2013,Zhu_PRD2025},
configuration entropy~\cite{Zhao_PLB2023,Braga_PRD2022}, heavy
quark-antiquark
potentials~\cite{Chen_JHEP2024,Zhou_PRD2020a,Zhou_PRD2020b,Zhang_NPB2023,Tahery_JHEP2023,Zhang_JHEP2011,Chu_JHEP2009,Hou_JHEP2008,Chen_EPJC2024}, and the effective running coupling of a heavy-quark pair and the
jet quenching parameter in a rotating medium~\cite{Zhu_PRD2026}.

In noncentral heavy-ion collisions, the incoming nuclei carry a
large orbital angular momentum, part of which can be transferred
to the produced QGP and induce spin polarization of quarks or
hadrons. For spin-one vector mesons, spin alignment provides an
important observable for studying the influence of the medium on
their spin states. The phenomenon was predicted in
Refs.~\cite{Liang_PRL2005,Liang_PLB2005}, and related spin
measurements have since been reported by the
STAR~\cite{Adamczyk_Nature2017,Adam_PRC2018,Adam_PRL2019,Abdallah_Nature2023}
and ALICE~\cite{Acharya_PRL2020,Acharya_PLB2021,Acharya_PRL2023}
Collaborations. STAR finds that the spin-alignment parameter
\(\rho_{00}\) of the \(\phi\) meson deviates from the unaligned
value of \(1/3\), with the magnitude of the deviation increasing
as the collision energy decreases~\cite{Abdallah_Nature2023}.
Taken together with the ALICE measurements, the data suggest that,
under the respective kinematic conditions and definitions of the
quantization axis, \(\phi\) and \(J/\psi\) mesons tend to exhibit
longitudinal and transverse polarization, respectively. These
preferences should be interpreted in the corresponding polarization
frames, and their dynamical origin remains unsettled. In particular,
the early-stage QGP is far from equilibrium, and its rotational and
vortical motion is time dependent, whereas many existing holographic
calculations employ static or stationary backgrounds. It is therefore
necessary to clarify the polarization patterns obtained within these
approximations and their range of applicability.

The spin state of a vector meson is described by a \(3\times3\)
spin density matrix \(\rho_{\lambda_1\lambda_2}\), where
\(\lambda_1,\lambda_2=0,\pm1\) denote spin projections along a
chosen quantization axis. The spin polarization, proportional to
\(\rho_{11}-\rho_{-1,-1}\), is not directly accessible from the
angular distributions of the parity-conserving decays considered here alone. The
spin-alignment parameter \(\rho_{00}\), however, can be extracted
from the polar angle distribution and represents the probability
for the meson to occupy the state with spin projection
\(\lambda=0\)~\cite{Sheng_PRD2024}. Experimentally, the relevant
channels are strong \(P\)-wave decays such as \(\phi\rightarrow
K^+K^-\)~\cite{Liang_PLB2005,Abdallah_Nature2023,Acharya_PRL2020,Yang_PRC2018}
and dilepton decays such as
\(J/\psi\rightarrow\mu^+\mu^-\)~\cite{Acharya_PLB2021,Acharya_PRL2023,Schilling_NPB1970,Faccioli_EPJC2010}.
For the latter, the polar angle distribution is conventionally
parametrized by \(\lambda_\theta\) as~\cite{Faccioli_EPJC2010}
\begin{equation}
W(\theta)\propto\frac{1}{3+\lambda_\theta}\left(1+\lambda_\theta\cos^2\theta\right).
\end{equation}
Here, \(\theta\) is the angle between the momentum of one decay
product and the chosen quantization axis in the meson rest frame.
Neglecting the lepton mass, the relation between
\(\lambda_\theta\) and \(\rho_{00}\)
is~\cite{Sheng_PRD2024,Zhao_JHEP2024}
\begin{equation}
\lambda_\theta=\frac{1-3\rho_{00}}{1+\rho_{00}}.
\end{equation}
The limiting cases \(\lambda_\theta=-1\) (\(\rho_{00}=1\)) and
\(\lambda_\theta=1\) (\(\rho_{00}=0\)) correspond to purely
longitudinal and purely transverse polarization, respectively.

Within holography, Sheng et al.~\cite{Sheng_PRD2024} developed a
general framework for the spin alignment of flavorless vector mesons,
relating spin-resolved dilepton production rates to spectral functions
in a thermal medium. An instantaneous freeze-out prescription connects
the in-medium spectral properties to final-state meson observables.
Using \(J/\psi\) and \(\phi\) in a soft-wall model as examples,
they studied spin alignment induced by the mesons' motion relative
to the thermal medium. At \(T=0.15~\mathrm{GeV}\), the \(J/\psi\)
spectral functions retain pronounced resonance peaks, whereas the
\(\phi\) spectral functions show no pronounced narrow peaks; their
helicity-frame \(\rho_{00}\) values exhibit positive and negative
deviations from \(1/3\), respectively. For a magnetized medium,
Zhao et al.~\cite{Zhao_JHEP2024} employed a soft-wall holographic
model to study the \(J/\psi\) spectral functions and spin parameters
\((\lambda_\theta,\lambda_\phi,\lambda_{\theta\phi})\). For meson
momentum parallel to the magnetic field, they found a positive
helicity-frame \(\lambda_\theta^H\) at low momentum that becomes
negative as momentum increases. Their results for \(\lambda_\theta\)
and \(\lambda_\phi\) show qualitative agreement with experimental
data in the helicity and Collins--Soper frames~\cite{Zhao_JHEP2024}.
Recently, we have also studied the spectral functions and spin
alignment of \(J/\psi\) and \(\phi\) at finite density in a
soft-wall holographic model~\cite{Wang_arXiv2026}.

Vector-meson spin alignment in rotating media has been studied within
the Nambu--Jona-Lasinio (NJL) model and its Polyakov-loop
extension~\cite{Wei_CPC2023,Sun_PRD2024}, as well as holographic
models. Chen et al.~\cite{Chen_arXiv2026} incorporated finite
temperature, baryon chemical potential, and angular velocity into
an anisotropic Einstein--Maxwell--dilaton (EMD) background and used
a four-flavor soft-wall model to study strange and charm vector
mesons. They attributed the flavor-dependent alignment patterns to
the different locations of the vacuum mass shell relative to the
in-medium longitudinal and transverse spectral peaks. Within the
parameter range considered, the spin alignment was insensitive to
baryon chemical potential and only weakly dependent on angular
velocity. Ahmed et al.~\cite{Ahmed_PRD2025} used a four-flavor
soft-wall model with rotation introduced through a \(U(1)\) gauge
field and induced gluon polarization described by an angular-velocity-dependent
dilaton field. They investigated the spectral functions and global
spin alignment of \(\rho\), \(\phi\), and \(J/\psi\) in the event
plane frame. Rotation increased the azimuthally averaged
\(\rho_{00}\) of \(\phi\) at higher transverse momentum, whereas
the \(J/\psi\) spin alignment remained largely insensitive to
temperature and rotation for \(p_T\lesssim5~\mathrm{GeV}\).
Ferreira~\cite{Ferreira_PRD2025} instead employed a rotating
background obtained by applying a Lorentz-like transformation to
the static AdS--Schwarzschild metric in cylindrical coordinates
and calculated the helicity-frame spin alignment of \(J/\psi\)
from spectral functions. At fixed angular velocity, increasing the
rotation radius improved the qualitative agreement of
\(\lambda_\theta\) with experimental data. However, the calculation
and data correspond to different rapidity ranges, so a comparison
under the same kinematic conditions is still needed. In related
work, Zhu et al.~\cite{Zhu_EPJC2024} used the black-brane
background obtained in the planar limit of a large five-dimensional
equal-spin Myers--Perry (MP) AdS black hole to study rotation
effects on heavy-quarkonium spectral functions and dissociation.

The five-dimensional equal-spin MP AdS black hole provides a
background with a well-defined gravitational origin for introducing
rotation. Its large-black-hole planar limit yields a uniformly
boosted black brane, simplifying the calculation while retaining
an explicit connection to the rotation parameter of the original
black hole. We employ this background to study the spectral
functions and spin alignment of \(J/\psi\) and \(\phi\) in a
common soft-wall framework, using the AdS--Schwarzschild black
brane recovered at \(a=0\) as the nonrotating baseline. Spin
alignment is determined from the relation between spin-resolved
dilepton production rates and in-medium spectral
functions~\cite{Sheng_PRD2024}. In the absence
of an explicit spin--vorticity coupling, we focus on how the
distinct spectral structures of the two mesons shape their
polarization responses. We systematically examine helicity-frame
spin alignment as a function of momentum magnitude and propagation
direction, and global spin alignment across transverse momenta,
rapidities, and azimuthal angles. This comparison provides a
reference for assessing the extent to which a uniformly boosted
background can describe vector-meson spin alignment and supplies
additional results for comparisons across backgrounds. Given the
homogeneous approximation, the results should not be interpreted
as isolated effects of local vorticity. Our discussion distinguishes
model-dependent numerical details from qualitative trends that
persist within the parameter range considered.

Section~\ref{sec:model} introduces the background and model;
Sec.~\ref{sec:dilepton} establishes the relation between dilepton
production rates and spin alignment; Sec.~\ref{sec:results}
presents the numerical results; and Sec.~\ref{sec:summary}
summarizes the work and discusses its limitations and future directions.

\section{Model setup}
\label{sec:model}
\subsection{Spinning Myers--Perry black-hole background}
To investigate the effects of rotation on the spectral functions
and spin alignment of vector mesons, we consider the
five-dimensional equal-spin MP AdS black
hole~\cite{Gibbons_CQG2005,Gibbons_PRL2004}. In the
large-black-hole planar limit, this geometry can be expressed as
an AdS--Schwarzschild black brane boosted along the \(z\)
direction. The metric is given by~\cite{Garbiso_JHEP2020}
\begin{equation}
        ds^2=\frac{1}{R^2\zeta^2}[
        \gamma^2(\beta^2-f)dt^2+dx^2+dy^2
        +\gamma^2(1-\beta^2f)dz^2
        +2\gamma^2\beta(1-f)dt\,dz]
        +\frac{R^2}{\zeta^2f}d\zeta^2,
    \label{eq:spinning_metric}
\end{equation}
where
\begin{equation}
    f(\zeta)=1-\frac{\zeta^4}{\zeta_h^4},\qquad
    \beta=\frac{a}{R},\qquad
    \gamma=\frac{1}{\sqrt{1-\beta^2}}.
\end{equation}
Here, \(R\) is the AdS curvature radius, \(a\) is the rotation
parameter, and \(\zeta\) is the holographic radial coordinate,
with \(\zeta_h\) denoting the horizon position. When \(a=0\), the
metric reduces to the AdS--Schwarzschild black brane. The Hawking
temperature of this background is given by~\cite{Garbiso_JHEP2020}
\begin{equation}
    T=\frac{\sqrt{R^2-a^2}}{\pi R^3\zeta_h}.
    \label{eq:spinning_temperature}
\end{equation}
Ref.~\cite{Garbiso_JHEP2020} suggests that the spinning black
brane may become unstable at \(a\approx0.75R\), so we restrict our
numerical calculations to \(0\leq a/R<0.75\), which also ensures a
positive temperature.

The frequency and temperature in the medium rest frame are given
by
\begin{equation}
    \bar\omega=\gamma(\omega+\beta p_z),\qquad
    \bar T=\gamma T=\frac{1}{\pi R^2\zeta_h}.
    \label{eq:rest_quantities}
\end{equation}
Here, \(p_z\) is the momentum component along the \(z\) direction.


\subsection{Holographic description of vector mesons}

We compute vector-current correlators in this background using
probe vector fields \(A_M\), neglecting their backreaction on the
geometry~\cite{Zhao_JHEP2024}. The action takes the Maxwell
form~\cite{Karch_PRD2006},
\begin{equation}
S_M=-\int d^4x\,d\zeta\,Q(\zeta)F_{MN}F^{MN}, \label{Sm}
\end{equation}
where \(Q(\zeta)=e^{-\phi(\zeta)}\sqrt{-g}/(4g_5^2)\) and
\(F_{MN}=\nabla_M A_N-\nabla_N A_M=\partial_M A_N-\partial_N
A_M\), with \(\nabla_M\) denoting the covariant derivative. Here
\(\phi(\zeta)\) is the dilaton profile and \(g_5\) is the gauge
coupling in five dimensions. The indices \(M,N\) label bulk
components in five dimensions, while \(\mu,\nu\) label spacetime
components in four dimensions. Varying the action gives
\begin{equation}
\partial_M\!\left[Q(\zeta)F^{MN}\right]=0.
\label{eom}
\end{equation}
We take the radial gauge \(A_\zeta=0\) and perform the Fourier
transform
\begin{equation}
A_\mu(x,\zeta)=\int\frac{d^4p}{(2\pi)^4} e^{-i\omega
t+i\boldsymbol p\cdot\boldsymbol x}A_\mu(p,\zeta),
\label{eq:vector_fourier}
\end{equation}
where \(p_\mu=(p_t,\boldsymbol p)=(-\omega,p_x,p_y,p_z)\). Using
Eqs.~\eqref{eom} and~\eqref{eq:vector_fourier}, the on-shell
action becomes~\cite{Zhao_JHEP2024}
\begin{equation}
S_M=-2\int\frac{d^4p}{(2\pi)^4} \left.
Q(\zeta)g^{\zeta\zeta}g^{\mu\nu} A_\mu(-p,\zeta)\partial_\zeta
A_\nu(p,\zeta) \right|_0^{\zeta_h}.
\label{eq:vector_boundary_action}
\end{equation}

Introducing the gauge-invariant electric
fields~\cite{Zhao_JHEP2024},
\begin{equation}
E_i(p,\zeta)\equiv-p_tA_i(p,\zeta)+p_iA_t(p,\zeta), \qquad
i=x,y,z, \label{eq:gauge_invariant_electric}
\end{equation}
we obtain
\begin{equation}
S_M=-2\int\frac{d^4p}{(2\pi)^4} \left.
Q(\zeta)\frac{g^{\zeta\zeta}}{p_t} g^{ik}G_k{}^j
E_i(-p,\zeta)\partial_\zeta E_j(p,\zeta) \right|_0^{\zeta_h}.
\label{eq:electric_boundary_action}
\end{equation}
For the present background with \(g^{tz}\neq0\), the radial
constraint \(g^{\mu\nu}p_\mu\partial_\zeta
A_\nu=0\)~\cite{Sheng_PRD2024} yields
\begin{equation}
g^{ik}G_k{}^j=-\frac{1}{p_t} \left[ g^{ij}-\frac{
\bigl(g^{i\alpha}p_\alpha\bigr)\bigl(g^{j\beta}p_\beta\bigr)
}{g^{\gamma\delta}p_\gamma p_\delta} \right].
\label{eq:electric_metric_combination}
\end{equation}
Here \(g^{\mu\nu}\) denotes the spacetime components in four
dimensions of the inverse bulk metric, including the off-diagonal
component \(g^{tz}\).

We choose three independent solutions \(\widetilde
E_i(e_j,\zeta,p)\) satisfying the ingoing-wave condition at the
horizon and normalized by \(\lim_{\zeta\to0}\widetilde
E_i(e_j,\zeta,p)=\delta_{ij}\) at the boundary. Here \(e_j\)
denotes the \(j\)th unit boundary source. The solution for an
arbitrary boundary source can then be written as
\begin{equation}
E_i(p,\zeta)=\widetilde E_i(e_j,\zeta,p)E_j^0(p), \qquad
(j=x,y,z). \label{eq:electric_source_decomposition}
\end{equation}
Here \(E_j^0(p)=\lim_{\zeta\to0}E_j(p,\zeta)\), and repeated
spatial indices are summed.

Substituting this decomposition into the boundary action gives
\begin{equation}
\begin{aligned}
S_M&=\int\frac{d^4p}{(2\pi)^4}E_l^0(-p)H^{ls}E_s^0(p),\\
H^{ls}&=-2Q(\zeta)\frac{g^{\zeta\zeta}}{p_t}g^{ik}G_k{}^j
\widetilde E_i(e_l,\zeta,-p)\partial_\zeta\widetilde
E_j(e_s,\zeta,p).
\end{aligned}
\label{eq:electric_boundary_kernel}
\end{equation}
The kernel \(H^{ls}=H^{ls}(\zeta,p)\) retains its radial
dependence; the radial boundary evaluation in the action is
implicit.

We return to the vector-field boundary sources
\(A_\mu^0(p)=\lim_{\zeta\to0}A_\mu(p,\zeta)\), which source the
dual current \(J^\mu\)~\cite{Sheng_PRD2024}. With
\(C_i^\mu=-p_t\delta_i^\mu+p_i\delta_t^\mu\), the sources satisfy
\(E_i^0(\pm p)=\pm C_i^\mu A_\mu^0(\pm p)\), giving
\begin{equation}
\begin{aligned}
S_M&=\int\frac{d^4p}{(2\pi)^4}A_\mu^0(-p)\mathcal F^{\mu\nu}A_\nu^0(p),\\
\mathcal F^{\mu\nu}&=-C_l^\mu H^{ls}C_s^\nu.
\end{aligned}
\label{eq:vector_boundary_kernel}
\end{equation}
Here \(\mathcal F^{\mu\nu}(\zeta,p)\) is the kernel of the
quadratic action in the vector-field boundary sources, defined by
the second line of
Eq.~\eqref{eq:vector_boundary_kernel}~\cite{Zhao_JHEP2024}. It is
distinct from the field strength \(F_{MN}\).

The Son--Starinets prescription~\cite{Son_JHEP2002} gives the
retarded correlator
\begin{equation}
D^{\mu\nu}(p)=2\lim_{\zeta\to0}\mathcal F^{\mu\nu}(\zeta,p).
\label{eq:retarded_correlator}
\end{equation}
The boundary limit in Eq.~\eqref{eq:retarded_correlator} is
understood after subtraction of real local contact terms, which do
not affect the spectral tensor~\cite{Son_JHEP2002}. The spectral
tensor is
\begin{equation}
\varrho^{\mu\nu}(p)=-\operatorname{Im}D^{\mu\nu}(p).
\label{eq:spectral_tensor}
\end{equation}
\subsection{Equations of motion}
To evaluate the correlation functions numerically, we solve
Eq.~\eqref{eom}~\cite{Zhao_JHEP2024}. The component form of
Eq.~\eqref{eom} reads

\begin{equation}
    \label{eq:eom_components}
    \begin{aligned}
        0={}&
        A_t''
        +\Bigl(-\frac{1}{\zeta}-\phi'-\gamma^2\beta^2\frac{f'}{f}\Bigr)A_t'
        +\gamma^2\beta\frac{f'}{f}A_z'
        -\frac{R^4}{f^2}\bigg\{
        f\bigl[p_x(\omega A_x+p_xA_t)+p_y(\omega A_y+p_yA_t)\bigr]
        \\
        &\qquad
        +\gamma^2\bigl[p_z(f-\beta^2)-\beta\omega(1-f)\bigr]
        (\omega A_z+p_zA_t)
        \bigg\},
        \\[0.9em]
        0={}&
        A_x''
        +\Bigl(-\frac{1}{\zeta}+\frac{f'}{f}-\phi'\Bigr)A_x'
        +\frac{R^4}{f^2}\bigg\{
        \gamma^2\bigl[\omega(1-\beta^2f)+\beta p_z(1-f)\bigr]
        (\omega A_x+p_xA_t)
        \\
        &\qquad
        +\gamma^2\bigl[p_z(f-\beta^2)-\beta\omega(1-f)\bigr]
        (p_xA_z-p_zA_x)
        +fp_y(p_xA_y-p_yA_x)
        \bigg\},
        \\[0.9em]
        0={}&
        A_y''
        +\Bigl(-\frac{1}{\zeta}+\frac{f'}{f}-\phi'\Bigr)A_y'
        +\frac{R^4}{f^2}\bigg\{
        \gamma^2\bigl[\omega(1-\beta^2f)+\beta p_z(1-f)\bigr]
        (\omega A_y+p_yA_t)
        \\
        &\qquad
        +\gamma^2\bigl[p_z(f-\beta^2)-\beta\omega(1-f)\bigr]
        (p_yA_z-p_zA_y)
        -fp_x(p_xA_y-p_yA_x)
        \bigg\},
        \\[0.9em]
        0={}&
        A_z''
        -\gamma^2\beta\frac{f'}{f}A_t'
        +\Bigl(-\frac{1}{\zeta}-\phi'+\gamma^2\frac{f'}{f}\Bigr)A_z'
        +\frac{R^4}{f^2}\bigg\{
        \gamma^2\bigl[\omega(1-\beta^2f)+\beta p_z(1-f)\bigr]
        (\omega A_z+p_zA_t)
        \\
        &\qquad
        -f\bigl[
        p_x(p_xA_z-p_zA_x)
        +p_y(p_yA_z-p_zA_y)
        \bigr]
        \bigg\},
        \\[0.9em]
        0={}&
        \gamma^2\bigl[\omega(1-\beta^2f)+\beta p_z(1-f)\bigr]A_t'
        +f(p_xA_x'+p_yA_y')
        +\gamma^2\bigl[p_z(f-\beta^2)-\beta\omega(1-f)\bigr]A_z'.
    \end{aligned}
\end{equation}

where the prime (\( ' \)) represents the derivative with respect
to \(\zeta\). These five equations describe the linear dynamics of
the probe gauge field in the boosted black-brane background. The
first four are equations of motion, and the last is the radial
constraint. The corresponding electric-field components,
\(E_j=\omega A_j+p_jA_t\) (with \(p_t=-\omega\)), obey
\begin{equation}
    \label{EOM_C}
    \begin{aligned}
        E_x''
        &+
        \left(
        -\frac{1}{\zeta}
        +\frac{f'}{f}
        -\phi'
        +p_x^2\Xi_T
        \right)E_x'
        +p_xp_y\Xi_T E_y'
        +p_x\Xi_P E_z'
        +\Theta E_x=0,
        \\
        E_y''
        &+
        \left(
        -\frac{1}{\zeta}
        +\frac{f'}{f}
        -\phi'
        +p_y^2\Xi_T
        \right)E_y'
        +p_xp_y\Xi_T E_x'
        +p_y\Xi_P E_z'
        +\Theta E_y=0,
        \\
        E_z''
        &+
        \left[
        -\frac{1}{\zeta}
        +\frac{f'}{f}
        -\phi'
        +(p_z+\beta\omega)\Xi_P
        \right]E_z'
        +(p_z+\beta\omega)\Xi_T
        \left(
        p_xE_x'
        +p_yE_y'
        \right)
        +\Theta E_z=0 .
    \end{aligned}
\end{equation}

\begin{equation}
    \begin{aligned}
        \Theta
        &=
        \frac{R^4}{f^2}
        \bigg\{
        \gamma^2(\omega+\beta p_z)^2
        -f
        \left[
        p_x^2+p_y^2
        +\gamma^2(p_z+\beta\omega)^2
        \right]
        \bigg\},
        \\[4pt]
        \Xi_T
        &=
        \frac{
            \gamma^2 f'(\omega+\beta p_z)
        }{
            \omega
            \bigg\{
            \gamma^2(\omega+\beta p_z)^2
            -f
            \left[
            p_x^2+p_y^2
            +\gamma^2(p_z+\beta\omega)^2
            \right]
            \bigg\}
        },
        \\[4pt]
        \Xi_P
        &=
        \frac{
            \gamma^2 f'
            \left[
            \omega(p_z+\beta\omega)
            -\beta(p_x^2+p_y^2)
            \right]
        }{
            \omega
            \bigg\{
            \gamma^2(\omega+\beta p_z)^2
            -f
            \left[
            p_x^2+p_y^2
            +\gamma^2(p_z+\beta\omega)^2
            \right]
            \bigg\}
        }.
    \end{aligned}
\end{equation}
Near the horizon, we impose an ingoing-wave condition. Thus
\(E(\zeta)\) can be written as
\begin{equation}
    \begin{aligned}
    E(\zeta)
    &=\left(1-\frac{\zeta}{\zeta_h}\right)^{-i\bar{\omega}/(4\pi\bar{T})}\psi(\zeta)\\
    &=\left(1-\frac{\zeta}{\zeta_h}\right)^{-i(\omega+\beta p_z)/(4\pi T)}\psi(\zeta),
    \end{aligned}
    \label{E_zeta}
\end{equation}
where \(\psi(\zeta)\) is regular at the horizon. Substituting this
factorization into Eq.~\eqref{EOM_C} and multiplying the resulting
system by \((\zeta/\zeta_h-1)^\nu\), with \(\nu\) the smallest
nonnegative integer that makes the coefficient matrices finite at
the horizon, gives~\cite{Sheng_PRD2024,Zhao_JHEP2024}
\begin{equation}
    \left(
    \widetilde{I}\,\partial_\zeta^2
    +\widetilde{K}\,\partial_\zeta
    +\widetilde{N}
    \right)\psi(\zeta)=0.
\end{equation}
Here \(\widetilde{I}\), \(\widetilde{K}\), and \(\widetilde{N}\)
are \(3\times3\) matrices, and \(\psi(\zeta)\) is a
three-component vector. Near the horizon, the regular solution
admits the expansion
\begin{equation}
    \psi(\zeta)=\psi_0+\sum_{n=1}^{\infty}a_n
    \left(\frac{\zeta}{\zeta_h}-1\right)^n. \label{taylor}
\end{equation}
where \(\psi_0\) represents the value of \(\psi(\zeta)\) at the
horizon and \(a_n\) is a \(3\times1\) matrix~\cite{Zhao_JHEP2024}.
Substituting the expansion into the matrix equation and equating
coefficients of powers of \((\zeta/\zeta_h-1)\) determines \(a_n\)
for a given \(\psi_0\)~\cite{Sheng_PRD2024,Zhao_JHEP2024}.
Numerically, we solve the longitudinal and transverse equations in
the medium rest frame and transform the resulting spectral tensor
back to the boosted frame using a Lorentz transformation. The
equations of motion, the ingoing boundary condition, and the
spectral tensor are listed above for completeness.

\section{Dilepton production through vector-meson decay}
\label{sec:dilepton}

We determine vector-meson spin alignment from the relation between
spin-resolved dilepton production rates and spectral functions in
the medium~\cite{Sheng_PRD2024,Zhao_JHEP2024}.

We consider dilepton production via the decay of a vector meson,
e.g., \(J/\psi\to l+\bar l\) or \(\phi\to l+\bar l\). For a
transition from an initial state \(i\) to a final state \(f\)
accompanied by a lepton pair \(l\bar l\), the \(S\)-matrix element
is~\cite{Gale_NPB1991}
\begin{equation}
    S_{fi}=\int d^4x\,d^4y\,\langle f,l\bar l|J^\mu(y)G_{\mu\nu}^R(x-y)J_l^\nu(x)|i\rangle.
    \label{Sfi}
\end{equation}

Equation~\eqref{Sfi} describes meson production at \(y\), followed
by propagation and decay into a lepton pair at \(x\). The retarded
propagator \(G_{\mu\nu}^R\) vanishes for \(x^0-y^0<0\), ensuring
causality.

In Eq.~\eqref{Sfi}, \(J^\mu\) is the current coupled to the vector
meson; for a QGP initial state, it represents the corresponding
quark current. The leptonic current \(J_l^\mu\) is
\begin{equation}
    J_l^\nu(x)=g_{Ml\bar l}\bar\psi_l(x)\Gamma^\nu\psi_l(x),
    \label{lepton_current}
\end{equation}
where \(\psi_l\) is the lepton field, \(g_{Ml\bar l}\) the
coupling strength, and \(\Gamma^\nu\) the effective interaction
vertex. We take
\(\Gamma^\nu\simeq\gamma^\nu\)~\cite{Sheng_PRD2024}.

The transition probability per unit spacetime volume is
\(R_{fi}\equiv |S_{fi}|^2/(TV)\), where \(TV\) is the spacetime
volume. After summing over final states and thermally averaging
over initial states, the differential dilepton production rate
\(n(x,p)\equiv dN/(d^4x\,d^4p)\) as a function of the total four-momentum \(p^\mu=(\omega,\boldsymbol p)\) is~\cite{Sheng_PRD2024}
\begin{align}
    n(x,p)={}&-\frac{2g_{Ml\bar l}^2}{3(2\pi)^5}\left(1-\frac{2m_l^2}{p^2}\right)\sqrt{1+\frac{4m_l^2}{p^2}}\,p^2n_B(x,\bar\omega) \notag\\
    &\times P_{\mu\nu}(p)G_A^{\mu\alpha}(p)\varrho_{\alpha\beta}(x,p)G_R^{\beta\nu}(p).
    \label{dilepton-rate}
\end{align}
Here the four-momentum transverse projector is defined by
\begin{equation}
P_{\mu\nu}(p)\equiv\eta_{\mu\nu}-\frac{p_\mu p_\nu}{p^2},
\end{equation}
which satisfies \(p^\mu P_{\mu\nu}=0\), with
\(p^2=-\omega^2+\boldsymbol p^2=-M^2\). Here \(m_l\) is the lepton
mass and \(G_A^{\mu\nu}\) is the advanced propagator. For the
present background, the Bose--Einstein distribution is
\begin{equation}
n_B(x,\bar\omega)=\frac{1}{e^{\bar\omega/\bar T}-1}
=\frac{1}{e^{(\omega+\beta p_z)/T}-1},
\end{equation}
where \(\bar\omega=-u_\mu p^\mu\) and \(\bar T\) are the frequency
and temperature in the medium rest frame, as defined in
Eq.~\eqref{eq:rest_quantities}. The spectral tensor in the medium
introduced in Sec.~\ref{sec:model} is
\begin{equation}
    \varrho_{\alpha\beta}(x,p)\equiv-\operatorname{Im}D_{\alpha\beta}(x,p),
    \label{spectral_function}
\end{equation}
with the retarded current-current correlator
\begin{equation}
    D^{\mu\nu}(x,p)\equiv-i\int d^4y\,\theta(y^0)\langle[J^\mu(y),J^\nu(0)]\rangle_{\bar T(x),u(x)}e^{-ip\cdot y}.
    \label{current_correlator}
\end{equation}
Here \(\langle O\rangle_{\bar T(x),u(x)}\) denotes the thermal
average~\cite{Sheng_PRD2024}. For the homogeneous background
considered here, \(u^\mu(x)=\gamma(1,0,0,-\beta)\) and \(\bar
T(x)=\bar T\).

We adopt an instantaneous freezeout approximation and neglect the
detailed freezeout kinematics~\cite{Sheng_PRD2024}. In
Eq.~\eqref{dilepton-rate}, \(\varrho_{\alpha\beta}\) describes the
spectral properties in the medium, whereas \(G_{R/A}^{\mu\nu}\)
describe effective propagation after freezeout, with the resonance
mass set to the vacuum meson mass. The spectral tensor is
sandwiched between these propagators as
\(G_A^{\mu\alpha}\varrho_{\alpha\beta}G_R^{\beta\nu}\).

Replacing \(G_{R/A}^{\mu\nu}\) by \(\eta^{\mu\nu}/p^2\) recovers
the standard virtual-photon dilepton
rate~\cite{Gale_NPB1991,Burnier_JHEP2009,McLerran_PRD1985,Weldon_PRD1990}.

Current conservation ensures that the production rate depends only
on the transverse part of the propagator in four dimensions. In
the positive-frequency resonance region, \(\omega>0\), we adopt
the Breit--Wigner form
\begin{equation}
G_{R/A}^{\mu\nu}(p)=-\frac{\eta^{\mu\nu}+p^\mu p^\nu/M^2}
{p^2+m_V^2\mp im_V\Gamma},\qquad M^2=-p^2,
\label{vector_propagator}
\end{equation}
where the upper (lower) sign corresponds to the retarded
(advanced) propagator. The numerator equals the transverse
projector \(P^{\mu\nu}(p)\) defined above and includes all three
physical polarization states, \(\lambda=0,\pm1\). Here \(m_V\) is
the vacuum meson mass and \(\Gamma\) is the width parameter of the
Breit--Wigner propagator.

The spectral function can be decomposed in a basis of polarization
vectors as
\begin{equation}
    \varrho^{\mu\nu}(x,p)=\sum_{\lambda,\lambda'=0,\pm1}v^\mu(\lambda,p)v^{*\nu}(\lambda',p)\widetilde{\varrho}_{\lambda\lambda'}(x,p). \label{varrho_uv}
\end{equation}
with the polarization vectors
\begin{equation}
    v^\mu(\lambda,p)=\left(\frac{\boldsymbol p\cdot\boldsymbol\epsilon_\lambda}{M},\boldsymbol\epsilon_\lambda+\frac{\boldsymbol p\cdot\boldsymbol\epsilon_\lambda}{M(\omega+M)}\boldsymbol p\right),
    \label{polarization_vector}
\end{equation}
which satisfy the orthonormality
\(\eta_{\mu\nu}v^\mu(\lambda,p)v^{*\nu}(\lambda',p)=\delta_{\lambda\lambda'}\)
and completeness
\begin{equation}
\sum_{\lambda}v^\mu(\lambda,p)v^{*\nu}(\lambda,p)
=\eta^{\mu\nu}-\frac{p^\mu p^\nu}{p^2} =\eta^{\mu\nu}+\frac{p^\mu
p^\nu}{M^2}.
\end{equation}
Here \(M\equiv\sqrt{\omega^2-\boldsymbol p^2}\) is the invariant
mass. The vectors \(\boldsymbol\epsilon_\lambda\) in three
dimensions specify the spin directions in the meson rest frame,
with \(\boldsymbol\epsilon_0\) defining the quantization direction
and \(\boldsymbol\epsilon_{\pm1}\) orthogonal to it.

The spin-space spectral components follow from the
projection~\cite{Zhao_JHEP2024}:
\[
\widetilde{\varrho}_{\lambda\lambda'}(x,p)
=v_\mu^*(\lambda,p)\,\varrho^{\mu\nu}(x,p)\,v_\nu(\lambda',p).
\]
For any chosen spin quantization axis, the dilepton rate separates
into spin-state contributions~\cite{Sheng_PRD2024}:
\begin{align}
    n_\lambda(x,p)={}&-\frac{2g_{Ml\bar l}^2}{3(2\pi)^5}\left(1-\frac{2m_l^2}{p^2}\right)\sqrt{1+\frac{4m_l^2}{p^2}} \notag\\
    &\times\frac{p^2n_B(x,\bar\omega)\widetilde{\varrho}_{\lambda\lambda}(x,p)}{(p^2+m_V^2)^2+m_V^2\Gamma^2}.
    \label{spin_resolved_rate}
\end{align}
Here, \(\widetilde{\varrho}_{\lambda\lambda'}\) denotes the full
spin-space spectral matrix, while its diagonal element
\(\widetilde{\varrho}_{\lambda\lambda}\) determines the production
rate in the spin channel \(\lambda\). When the quantization axis
is along \(\boldsymbol p\), the \(\lambda=0\) and \(\lambda=\pm1\)
states correspond to longitudinal and transverse polarization,
respectively. The total rate is
\(n(x,p)=\sum_{\lambda=0,\pm1}n_\lambda(x,p)\). For a vector-meson
resonance, the spin alignment is defined as the probability for
the zero-projection state~\cite{Sheng_PRD2024},
\begin{equation}
    \rho_{00}(x,\boldsymbol p)\equiv
    \frac{\displaystyle\int d\omega\,n_0(x,p)}
    {\displaystyle\sum_{\lambda=0,\pm1}\int d\omega\,n_\lambda(x,p)},
    \label{spin_alignment}
\end{equation}
Since each \(n_\lambda\) is nonnegative, \(0\leq\rho_{00}\leq1\).
For \(\Gamma\ll m_V\), provided the remaining factors vary slowly
across the resonance peak at \(-p^2=m_V^2\), the spin alignment
can be approximated by~\cite{Sheng_PRD2024}
\begin{equation}
\rho_{00}(x,\boldsymbol p)\simeq \left.
\frac{\widetilde{\varrho}_{00}(x,p)}
{\sum_{\lambda=0,\pm1}\widetilde{\varrho}_{\lambda\lambda}(x,p)}
\right|_{\omega=\sqrt{m_V^2+\boldsymbol p^2}}.
\label{spin_alignment_narrow_width}
\end{equation}
We use this narrow-width approximation for the numerical spin
alignment results, evaluating the spectral functions at \(M=m_V\).
For the \(\phi\) meson, the absence of a narrow resonance means
that Eq.~\eqref{spin_alignment_narrow_width} should be read as an
operational definition of the alignment at the vacuum mass rather
than as a controlled narrow-width limit. We use it for the
qualitative comparison and treat the resulting absolute values for
\(\phi\) as carrying an additional model uncertainty. A full
\(\omega\)-integral evaluation would require a more detailed
freezeout prescription and is left for future work.

\section{Numerical results}
\label{sec:results}

We set \(R=1\) and take \(a=0\), \(0.1\), \(0.2\), and \(0.3\). In
the helicity frame, the quantization axis lies along the
propagation direction. Longitudinal and transverse polarizations
correspond to \(\lambda=0\) and \(\lambda=\pm1\), respectively.

For global spin alignment, the quantization axis is \(\boldsymbol
e_0^y=(0,1,0)\). We choose \(z\) as the beam and rapidity axis and
parametrize the meson momentum as
\begin{equation}
\boldsymbol p=(p_T\cos\varphi,p_T\sin\varphi,m_T\sinh Y),
\end{equation}
where \(m_T=\sqrt{M^2+p_T^2}\), \(p_T\) is the transverse momentum
in the \(x\)--\(y\) plane, and \(Y\) and \(\varphi\) are the
rapidity and azimuthal angle, respectively.

The momentum and rapidity ranges \(p_T=1\)--\(10~\mathrm{GeV}\)
and \(Y=0\)--\(1\) are chosen to cover the kinematic ranges
typically probed in measurements of \(J/\psi\) and \(\phi\) spin
alignment at the Relativistic Heavy Ion Collider (RHIC) and the
Large Hadron Collider (LHC). The main calculations use
\(T=0.15\)--\(0.20~\mathrm{GeV}\); for the \(\phi\) meson, we
also consider \(T=0.10~\mathrm{GeV}\). These temperature choices
are motivated by previous soft-wall holographic studies of
vector-meson spectral functions and spin
alignment~\cite{Zhao_JHEP2024,Sheng_PRD2024}.
The bound \(a/R<0.75\) ensures background stability. Because
varying \(a\) at fixed Hawking temperature also changes the
medium-rest-frame temperature, the results below should be read
as a homogeneous rotating-medium reference calculation rather
than as isolated local-vorticity effects.
In particular, when \(a=0\), our background reduces to the
AdS--Schwarzschild black brane employed by Sheng et al., providing
a consistent nonrotating baseline for investigating the effects of
the spinning background.

\subsection{\(J/\psi\) meson}
We begin with the dimuon decay channel of the \(J/\psi\). The
vacuum mass is
\(m_{J/\psi}=3.096~\mathrm{GeV}\)~\cite{Navas_PRD2024}. The
dilaton profile in Eq.~\eqref{Sm} is chosen as
\(\phi(\zeta)=c_{J/\psi}\zeta^2\), with
\(c_{J/\psi}=m_{J/\psi}^2/4\simeq2.40~\mathrm{GeV}^2\)~\cite{Sheng_PRD2024,Zhao_JHEP2024}.

We first consider \(J/\psi\) mesons propagating along the \(z\)
direction, with \(\boldsymbol p=(0,0,p)\) and \(p\geq0\).

\begin{figure}[htbp]
    \centering
    \includegraphics[width=8cm]{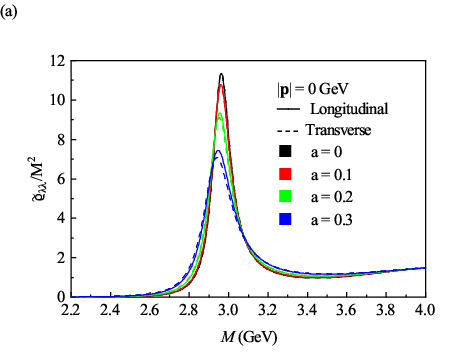}
    \includegraphics[width=8cm]{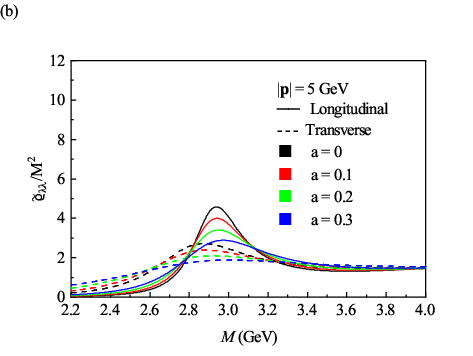}
    \includegraphics[width=8cm]{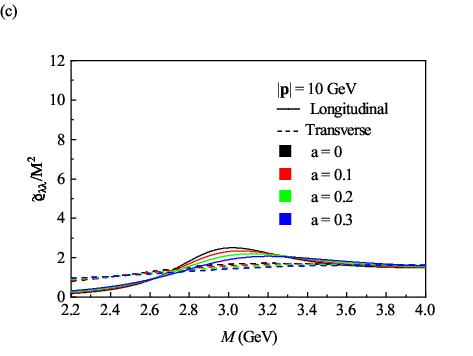}
    \caption{Spectral functions \(\tilde{\varrho}_{\lambda\lambda}/M^2\) of \(J/\psi\) mesons versus invariant mass \(M\) at \(T=0.15\) GeV for \(\boldsymbol p=(0,0,p)\). Panels (a) to (c) show \(|\boldsymbol p|=0\), \(5\), and \(10\) GeV. Solid and dashed curves denote longitudinal and transverse polarizations, respectively.}
    \label{fig:jpsi_spectral}
\end{figure}

Figure~\ref{fig:jpsi_spectral} shows the \(J/\psi\) spectral
functions versus invariant mass \(M\) at \(T=0.15~\mathrm{GeV}\)
for different momenta and rotation parameters. At fixed momentum,
increasing \(a\) generally lowers and broadens the resonance
peaks, with slight changes in their positions. The differences
between spectral functions at different \(a\) are more pronounced
at low momentum and become smaller at \(|\boldsymbol
p|=10~\mathrm{GeV}\).

The momentum dependence is as follows. For the \(a=0\) baseline at
zero momentum, the three spin states are degenerate and the
spectral functions exhibit a narrow peak near
\(M\simeq3.0~\mathrm{GeV}\), as shown in
Fig.~\ref{fig:jpsi_spectral}(a). Nonzero \(a\) or finite momentum
separates the longitudinal and transverse spectral functions. As
momentum increases from \(0\) to \(5\) and \(10~\mathrm{GeV}\),
the peaks become lower and broader, as shown in
Figs.~\ref{fig:jpsi_spectral}(b) and (c). At finite momentum, the
transverse spectral functions are larger at lower invariant
masses, whereas the longitudinal ones are larger near the
resonance peaks; the two approach each other at higher invariant
masses.

\begin{figure}[htbp]
    \centering
    \includegraphics[width=8cm]{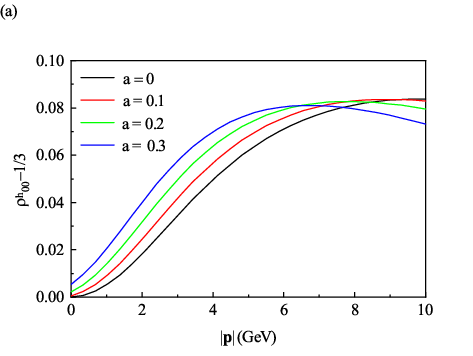}
    \includegraphics[width=8cm]{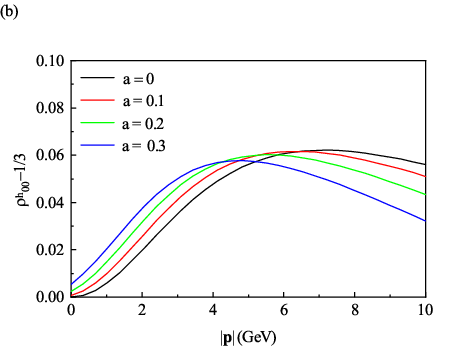}
    \includegraphics[width=8cm]{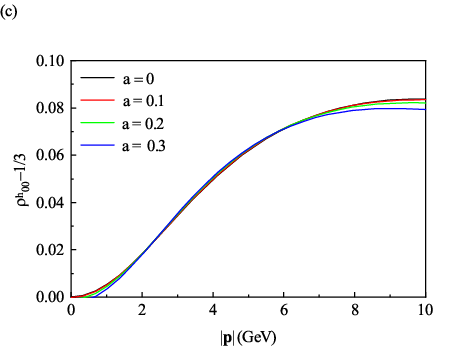}
    \includegraphics[width=8cm]{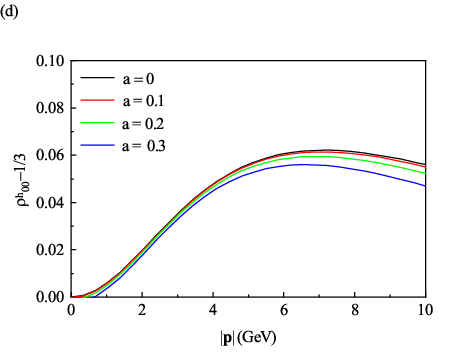}
    \caption{Helicity-frame spin alignment \(\rho_{00}^{h}-1/3\) of \(J/\psi\) mesons versus \(|\boldsymbol p|\). Panels (a,b) and (c,d) show propagation along the \(z\) and \(x\) directions, respectively; \(T=0.15\) GeV in (a,c) and \(T=0.20\) GeV in (b,d). For propagation along \(x\) at finite \(a\), small negative deviations occur at very low momentum and are not shown.}
    \label{fig:jpsi_helicity}
\end{figure}

The helicity-frame spin alignment is obtained from the spectral
functions using Eq.~\eqref{spin_alignment_narrow_width}.
Figure~\ref{fig:jpsi_helicity} shows its momentum dependence for
propagation along \(z\) and \(x\) at different temperatures and
rotation parameters. The deviation \(\rho_{00}^{h}-1/3\) is
positive over most of the momentum range, corresponding to a
negative \(\lambda_\theta\). This result is qualitatively
different from the polarization tendency observed experimentally;
the difference may be related to the momentum range
considered~\cite{Sheng_PRD2024}. The deviation
vanishes at \(a=0\) and \(|\boldsymbol p|=0\). For finite \(a\),
the numerical results for propagation along \(x\) also contain
small negative deviations at very low momentum.

For propagation along \(z\), as shown in
Figs.~\ref{fig:jpsi_helicity}(a) and (b), the deviation at
\(T=0.15~\mathrm{GeV}\) initially increases with momentum and
approaches a broad maximum at larger momenta, after which some
curves decrease. Increasing \(a\) enhances the positive deviation
at low and intermediate momenta. At high momentum, the curves
cross and this effect can change to suppression; their ordering
depends on momentum and the values of \(a\).

For propagation along \(x\), with \(\boldsymbol p=(p,0,0)\), as
shown in Figs.~\ref{fig:jpsi_helicity}(c) and (d), the results for
different \(a\) at \(T=0.15~\mathrm{GeV}\) are closer than those
for propagation along \(z\). The deviation varies more slowly at
high momentum, where increasing \(a\) slightly reduces the
positive deviation.

When the temperature increases to \(T=0.20~\mathrm{GeV}\), the
positive deviations for both propagation directions become smaller
and decrease after reaching maxima, which generally occur at lower
momenta. For propagation along \(z\), increasing \(a\) still
produces an enhancement at low and intermediate momenta, whereas
the suppression at high momentum becomes more pronounced. For
propagation along \(x\), the differences between results at
different \(a\) remain small.

We next examine the global spin alignment. In the isotropic
background at \(a=0\), the two transverse states have equal
spectral weights in the helicity basis. Defining
\(\Sigma(x,p)=\sum_{\lambda=0,\pm1}\widetilde{\varrho}_{\lambda\lambda}^{\,h}(x,p)\)
and \(c=\boldsymbol e_0^y\cdot\boldsymbol e_0^h\), we obtain

\begin{align}
    \widetilde{\varrho}_{00}^{\,y}(x,p)
    ={}&
    \frac{1}{2}
    \{
    \Sigma(x,p)-\widetilde{\varrho}_{00}^{\,h}(x,p)
    +
    c^2
    [
    3\widetilde{\varrho}_{00}^{\,h}(x,p)-\Sigma(x,p)
    ]
    \},
    \nonumber\\
    \sum_{\lambda}
    \widetilde{\varrho}_{\lambda\lambda}^{\,y}(x,p)
    ={}&
    \sum_{\lambda}
    \widetilde{\varrho}_{\lambda\lambda}^{\,h}(x,p).
\end{align}

Using the spectral ratio in
Eq.~\eqref{spin_alignment_narrow_width}, the above relations give

\begin{equation}
    \rho_{00}^{\,y}(x,p)
    =
    \frac{1}{3}
    +
    \frac{3\rho_{00}^{\,h}(x,p)-1}{3|\mathbf{p}|^2}
    \left(
    p_y^2-\frac{p_x^2+p_z^2}{2}
    \right).
\end{equation}

\begin{figure}[htbp]
    \centering
    \includegraphics[width=8cm]{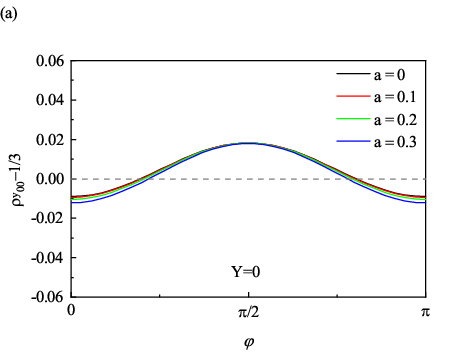}
    \includegraphics[width=8cm]{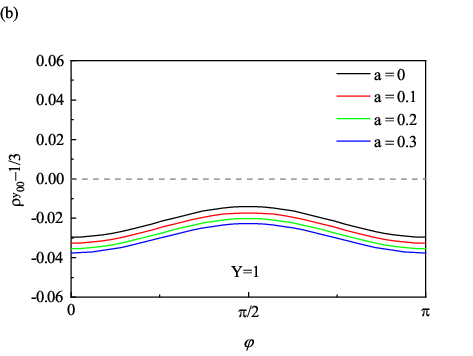}
    \caption{Global spin alignment \(\rho_{00}^{y}-1/3\) of \(J/\psi\) mesons versus \(\varphi\) at \(T=0.15\) GeV and \(p_T=2\) GeV. Panels (a) and (b) show \(Y=0\) and \(Y=1\), respectively.}
    \label{fig:jpsi_azimuth}
\end{figure}

At \(a=0\) and \(Y=0\), \(c^2=\sin^2\varphi\). A positive
(negative) deviation in the helicity frame therefore corresponds
to a maximum (minimum) at \(\varphi=\pi/2\), explaining the
opposite shapes of the nonrotating baseline curves in
Figs.~\ref{fig:jpsi_azimuth} and \ref{fig:phi_azimuth}. The
angular projection structure is also similar to the corresponding
structure for vector-meson spin alignment in the quark-coalescence
model~\cite{Sheng_PRC2023}. For finite \(a\) or general
kinematics, we use the polarization projection of the full
spectral tensor~\cite{Sheng_PRD2024}:

\begin{equation}
\widetilde{\varrho}_{00}^{\,y}(x,p)
=v_\mu^*(0_y,p)\,\varrho^{\mu\nu}(x,p)\,v_\nu(0_y,p).
\end{equation}

The corresponding spin alignment is calculated using
Eq.~\eqref{spin_alignment_narrow_width}.

Figure~\ref{fig:jpsi_azimuth} shows the global spin alignment
\(\rho_{00}^{y}-1/3\) versus \(\varphi\) at
\(T=0.15~\mathrm{GeV}\) and \(p_T=2~\mathrm{GeV}\). At midrapidity
\(Y=0\), shown in Fig.~\ref{fig:jpsi_azimuth}(a), the distribution
is approximately symmetric about \(\varphi=\pi/2\):
\(\rho_{00}^{y}\) is below \(1/3\) near \(\varphi=0\) and \(\pi\),
and reaches a maximum above \(1/3\) near \(\varphi=\pi/2\). As
\(a\) increases, the results near the center remain nearly
unchanged, while the negative deviations near the endpoints
increase slightly.

At larger rapidity, \(Y=1\), shown in
Fig.~\ref{fig:jpsi_azimuth}(b), the distribution still reaches its
maximum near \(\varphi=\pi/2\), but \(\rho_{00}^{y}\) remains
below \(1/3\) over the entire azimuthal range. Increasing \(a\)
increases the negative deviation of \(\rho_{00}^{y}\) from
\(1/3\), with the largest deviation at \(a=0.3\).

The dependence of the global spin alignment on the rotation
parameter is therefore more pronounced at \(Y=1\) than at \(Y=0\).
Within the parameter range considered, the basic shapes of the
azimuthal distributions remain unchanged at both rapidities.

\begin{figure}[htbp]
    \centering
    \includegraphics[width=8cm]{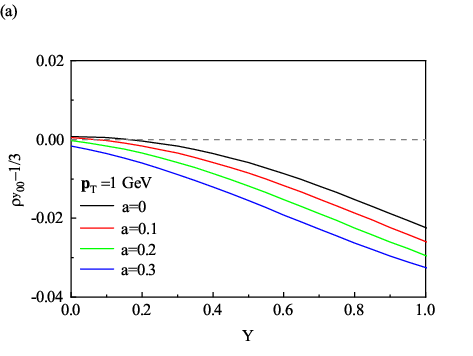}
    \includegraphics[width=8cm]{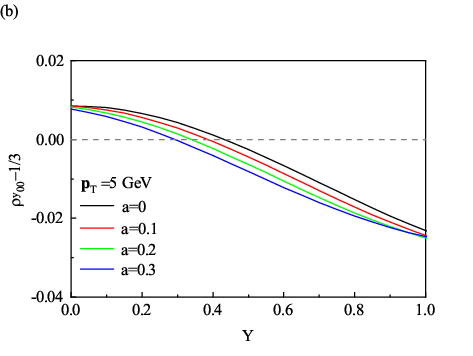}
    \includegraphics[width=8cm]{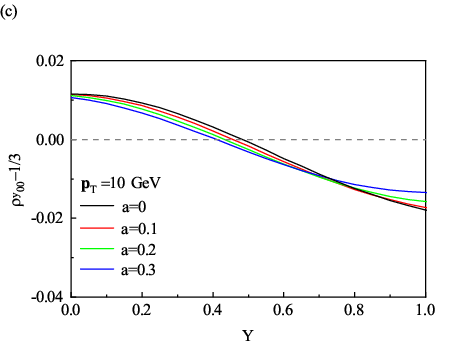}
    \caption{Azimuthally averaged global spin alignment \(\rho_{00}^{y}-1/3\) of \(J/\psi\) mesons versus \(Y\) at \(T=0.15\) GeV. Panels (a) to (c) show \(p_T=1\), \(5\), and \(10\) GeV, respectively.}
    \label{fig:jpsi_rapidity}
\end{figure}

After averaging over \(\varphi\), Fig.~\ref{fig:jpsi_rapidity}
shows the global spin alignment of the \(J/\psi\) versus rapidity
\(Y\) for \(p_T=1\), \(5\), and \(10~\mathrm{GeV}\). We use the
weight \(1+2v_2\cos(2\varphi)\) with
\(v_2=0.15\)~\cite{Sheng_PRD2024}. Over the rapidity range shown,
\(\rho_{00}^{y}-1/3\) decreases with increasing \(Y\). At
\(p_T=1~\mathrm{GeV}\), the deviation decreases from a value close
to zero near midrapidity to negative values; at \(p_T=5\) and
\(10~\mathrm{GeV}\), it changes from positive to negative. The
deviation near midrapidity increases with \(p_T\), but this
ordering does not hold over the entire rapidity range.

The rotation parameter affects this rapidity dependence. As shown
in Fig.~\ref{fig:jpsi_rapidity}(a), increasing \(a\) decreases
\(\rho_{00}^{y}\) at \(p_T=1~\mathrm{GeV}\). The same trend is
observed at \(p_T=5~\mathrm{GeV}\) in
Fig.~\ref{fig:jpsi_rapidity}(b), although the results for
different \(a\) approach each other at larger rapidity. At
\(p_T=10~\mathrm{GeV}\), the differences between results at
different \(a\) are smaller, and the curves cross at larger
rapidity, as shown in Fig.~\ref{fig:jpsi_rapidity}(c); their
ordering therefore changes with rapidity.


\subsection{\(\phi\) meson}

For the \(\phi\) meson, we use the dilaton profile
\(\phi(\zeta)=c_\phi\zeta^2\), with
\(c_\phi=m_\phi^2/4\simeq0.26~\mathrm{GeV}^2\) and
\(m_\phi=1.02~\mathrm{GeV}\)~\cite{Sheng_PRD2024}. We consider the
decay \(\phi\to\mu^+\mu^-\) and assume that spin alignment is
independent of, or only weakly dependent on, the decay
channel~\cite{Sheng_PRD2024}.

We first consider \(\phi\) mesons propagating along the \(z\)
direction, with \(\boldsymbol p=(0,0,p)\).

\begin{figure}[htbp]
    \centering
    \includegraphics[width=8cm]{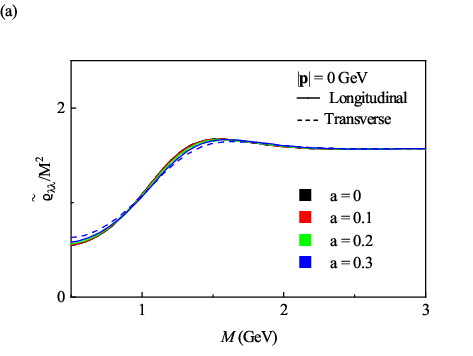}
    \includegraphics[width=8cm]{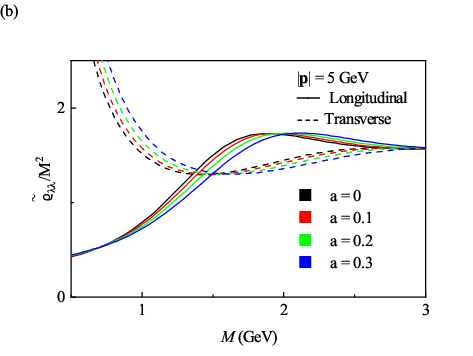}
    \includegraphics[width=8cm]{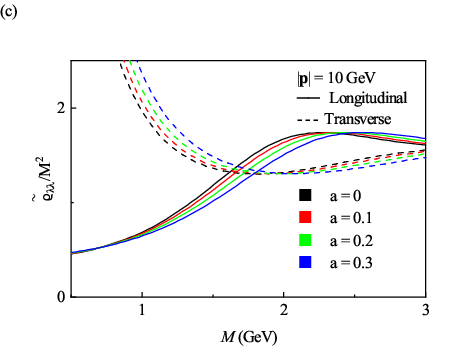}
    \caption{Spectral functions \(\tilde{\varrho}_{\lambda\lambda}/M^2\) of \(\phi\) mesons versus invariant mass \(M\) at \(T=0.15\) GeV for \(\boldsymbol p=(0,0,p)\). Panels (a) to (c) show \(|\boldsymbol p|=0\), \(5\), and \(10\) GeV. Solid and dashed curves denote longitudinal and transverse polarizations, respectively.}
    \label{fig:phi_spectral}
\end{figure}

Figure~\ref{fig:phi_spectral} presents the longitudinal and
transverse spectral functions of the \(\phi\) meson versus
invariant mass \(M\) at \(T=0.15\) GeV. In contrast to the
\(J/\psi\) meson shown in Fig.~\ref{fig:jpsi_spectral}, the
\(\phi\) spectral functions exhibit no pronounced resonance peaks
near the vacuum mass, but instead display relatively broad
spectral structures. This indicates that the bound-state
characteristics of the \(\phi\) meson are significantly weakened
in the medium at this temperature.

At \(|\boldsymbol p|=0\), the longitudinal and transverse spectral
functions nearly coincide and depend only weakly on \(a\). At
\(|\boldsymbol p|=5\) and \(10~\mathrm{GeV}\), the difference
between the two polarizations becomes larger: the transverse
contributions dominate at low invariant masses, while the
longitudinal contributions increase with \(M\) and exceed the
transverse contributions over part of the mass range. The rotation
parameter also affects the spectral distributions in different
polarization channels. At these two momenta, increasing \(a\)
enhances the transverse spectral functions at low invariant masses
and shifts the broad longitudinal structures toward higher
invariant masses.

Within the instantaneous freezeout approximation adopted in
Sec.~\ref{sec:dilepton}, the vacuum propagators suppress
contributions far from \(m_\phi\). Thus, even without a pronounced
narrow peak in the spectral functions in the medium, the
probability distribution of \(\phi\) mesons after freezeout can
develop resonance peaks near the vacuum mass and contribute to the
dilepton production rate~\cite{Sheng_PRD2024}.

\begin{figure}[htbp]
    \centering
    \includegraphics[width=8cm]{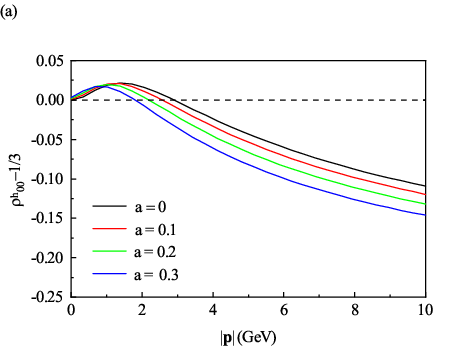}
    \includegraphics[width=8cm]{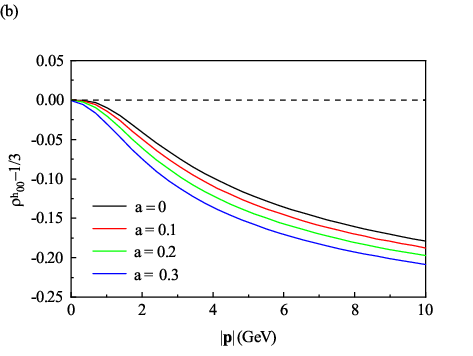}
    \includegraphics[width=8cm]{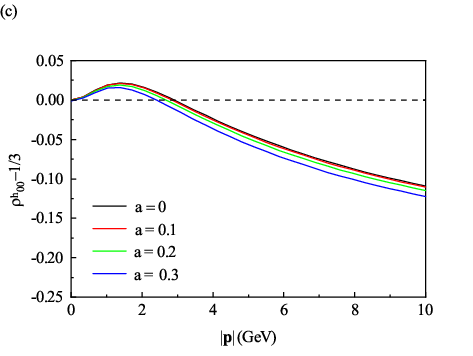}
    \includegraphics[width=8cm]{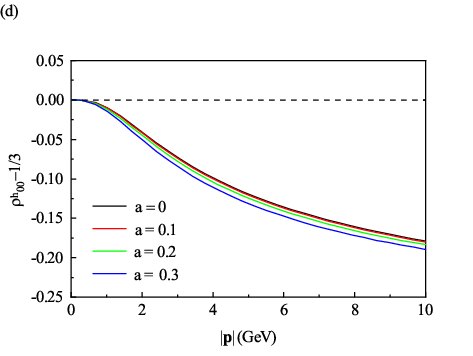}
    \caption{Helicity-frame spin alignment \(\rho_{00}^{h}-1/3\) of \(\phi\) mesons versus \(|\boldsymbol p|\). Panels (a,b) and (c,d) show propagation along the \(z\) and \(x\) directions, respectively; \(T=0.10\) GeV in (a,c) and \(T=0.15\) GeV in (b,d).}
    \label{fig:phi_helicity}
\end{figure}

Figure~\ref{fig:phi_helicity} shows the helicity-frame spin
alignment \(\rho_{00}^{h}-1/3\) of the \(\phi\) meson versus
\(|\boldsymbol p|\) at \(T=0.10\) and \(0.15\) GeV for different
rotation parameters. For propagation along \(z\), shown in
Figs.~\ref{fig:phi_helicity}(a) and (b), increasing \(a\) at fixed
temperature generally reduces \(\rho_{00}^{h}\) in the
intermediate and high-momentum regions.

For propagation along \(x\), with \(\boldsymbol p=(p,0,0)\), shown
in Figs.~\ref{fig:phi_helicity}(c) and (d), the effect of \(a\) is
weaker than for propagation along \(z\); the results for different
values of \(a\) are especially close at \(T=0.15~\mathrm{GeV}\).

The temperature changes the form of the momentum dependence. At
\(T=0.10~\mathrm{GeV}\), both propagation directions show a
positive deviation that reaches a maximum and then becomes
negative at larger momenta. At \(T=0.15~\mathrm{GeV}\), the
deviation is negative over most of the momentum range, and its
absolute value increases with momentum. At the higher temperature,
the predominantly negative deviations correspond to a preference
for transverse polarization.

\begin{figure}[htbp]
    \centering
    \includegraphics[width=8cm]{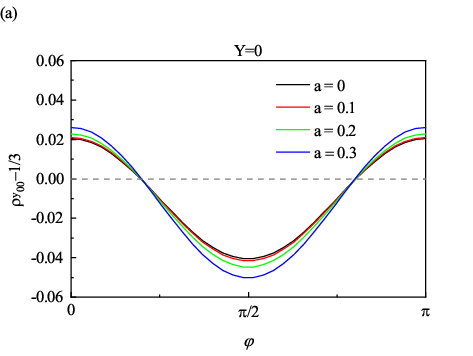}
    \includegraphics[width=8cm]{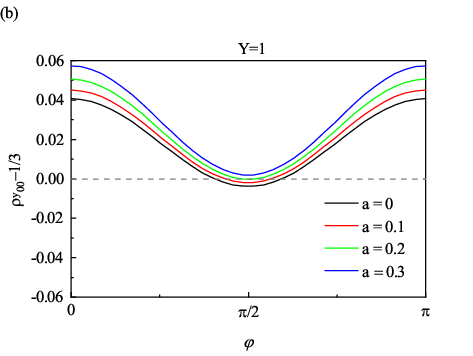}
    \caption{Global spin alignment \(\rho_{00}^{y}-1/3\) of \(\phi\) mesons versus \(\varphi\) at \(T=0.15\) GeV and \(p_T=2\) GeV. Panels (a) and (b) show \(Y=0\) and \(Y=1\), respectively.}
    \label{fig:phi_azimuth}
\end{figure}

Figure~\ref{fig:phi_azimuth} shows the global spin alignment
\(\rho_{00}^{y}-1/3\) of the \(\phi\) meson versus \(\varphi\) at
\(p_T=2\) GeV and \(T=0.15\) GeV. For both \(Y=0\) and \(Y=1\),
the distribution is symmetric about \(\varphi=\pi/2\) and reaches
its minimum at this angle.

At midrapidity \(Y=0\), shown in Fig.~\ref{fig:phi_azimuth}(a),
the deviation \(\rho_{00}^{y}-1/3\) is positive near \(\varphi=0\)
and \(\pi\) and negative near \(\varphi=\pi/2\), opposite to the
midrapidity pattern of the \(J/\psi\) in
Fig.~\ref{fig:jpsi_azimuth}. Increasing \(a\) enhances both the
positive deviations at the endpoints and the negative deviation at
the center, increasing the azimuthal modulation.

At \(Y=1\), shown in Fig.~\ref{fig:phi_azimuth}(b),
\(\rho_{00}^{y}\) is generally larger than at \(Y=0\) and
approaches \(1/3\) near \(\varphi=\pi/2\). Increasing \(a\)
further increases \(\rho_{00}^{y}\), so that the value at the
center slightly exceeds \(1/3\) for larger \(a\). The shape of the
distribution, with a minimum at the center and maxima at the
endpoints, remains largely unchanged.

\begin{figure}[htbp]
    \centering
    \includegraphics[width=8cm]{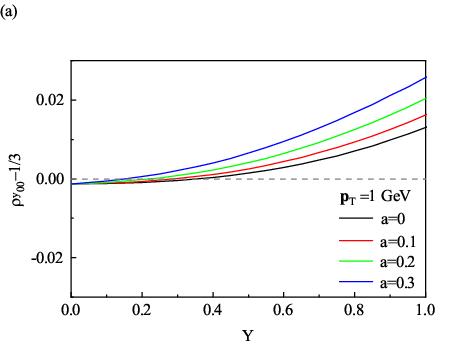}
    \includegraphics[width=8cm]{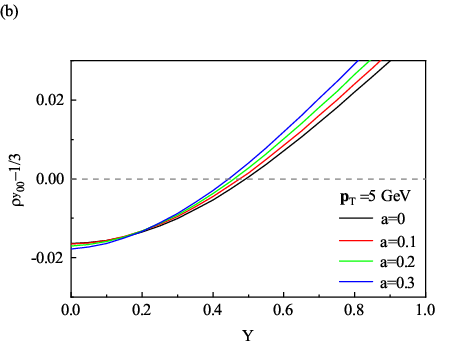}
    \includegraphics[width=8cm]{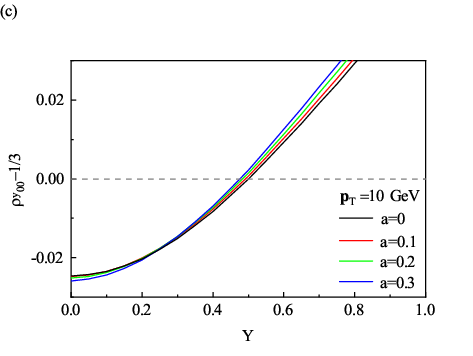}
    \caption{Azimuthally averaged global spin alignment \(\rho_{00}^{y}-1/3\) of \(\phi\) mesons versus \(Y\) at \(T=0.15\) GeV. Panels (a) to (c) show \(p_T=1\), \(5\), and \(10\) GeV, respectively.}
    \label{fig:phi_rapidity}
\end{figure}

After the same azimuthal averaging, Fig.~\ref{fig:phi_rapidity}
shows the global spin alignment of the \(\phi\) meson versus
rapidity \(Y\) for \(p_T=1\), \(5\), and \(10\) GeV in panels (a)
to (c), respectively. Within the range considered,
\(\rho_{00}^{y}-1/3\) increases with \(Y\), opposite to the
decreasing trend of the \(J/\psi\) in
Fig.~\ref{fig:jpsi_rapidity}. At \(p_T=1~\mathrm{GeV}\), the
deviation is small near midrapidity and gradually becomes
positive. With the adopted weighting at \(v_2=0.15\), the
transition from negative to positive deviation occurs at
approximately \(Y=0.5\) for \(p_T=5\) and \(10~\mathrm{GeV}\), and
the rapidity dependence becomes more pronounced as \(p_T\)
increases.

The rotation parameter also affects the rapidity dependence. At
\(p_T=1~\mathrm{GeV}\), increasing \(a\) increases
\(\rho_{00}^{y}\) at finite rapidity, as shown in
Fig.~\ref{fig:phi_rapidity}(a). At \(p_T=5~\mathrm{GeV}\), it
slightly reduces \(\rho_{00}^{y}\) at lower rapidities but
increases it at larger rapidities, as shown in
Fig.~\ref{fig:phi_rapidity}(b). At \(p_T=10~\mathrm{GeV}\), the
results for different values of \(a\) remain close, as shown in
Fig.~\ref{fig:phi_rapidity}(c).

The magnitude of the spin-alignment deviation reaches the order of
\(\mathcal{O}(10^{-2})\), comparable to that observed in the STAR
experiment~\cite{Abdallah_Nature2023}. We emphasize that the
numerical positions of the peaks, the precise crossing momenta,
and the detailed dependence on \(a\) at fixed Hawking temperature
are model dependent. The qualitative features that are stable
within the present soft-wall model class are the opposite signs of
the helicity-frame deviations for \(J/\psi\) and \(\phi\) over
most of the momentum range at \(T=0.15~\mathrm{GeV}\), the
stronger effect of rotation for propagation along \(z\) than along
\(x\), and the opposite azimuthal and rapidity trends of the
global spin alignment for the two mesons.

\clearpage
\section{Summary and outlook}
\label{sec:summary}

We have investigated the effects of the rotation parameter and
meson momentum on the spectral functions and spin alignment of
\(J/\psi\) and \(\phi\) mesons using a soft-wall holographic model
in the planar limit of a five-dimensional equal-spin MP AdS black
hole. Compared with the nonrotating study of
Ref.~\cite{Sheng_PRD2024} and the spectral-function study of
Ref.~\cite{Zhu_EPJC2024}, the present work provides a unified spin
alignment analysis for both charmonium and light vector mesons in
the same rotating background, and quantifies the momentum,
rapidity, and azimuthal dependencies that are relevant for
experimental comparisons. The equations of motion, the ingoing
boundary condition, and the spectral tensor are given in
Sec.~\ref{sec:model} for reference.

At \(T=0.15~\mathrm{GeV}\), the \(J/\psi\) spectral functions
exhibit significant resonance peaks at low momentum, indicating
that \(J/\psi\) can exist as a quasistable meson in the medium.
The peaks become lower and broader with increasing momentum. For
the \(J/\psi\), increasing the rotation parameter generally lowers
and broadens the resonance peaks. At the same temperature, the
\(\phi\) spectral functions show no significant narrow resonance
peak near the vacuum mass and can be interpreted as probability
distributions of unstable \(s\bar{s}\) pairs in the thermal
medium. For \(\phi\) mesons propagating along \(z\) with
\(|\boldsymbol p|=5\) and \(10~\mathrm{GeV}\), increasing \(a\)
enhances the transverse spectral functions at low invariant masses
and shifts the broad longitudinal structures toward higher
invariant masses.

We have also examined the helicity-frame spin alignment for mesons
propagating along the \(z\) and \(x\) directions. At
\(T=0.15~\mathrm{GeV}\), over most of the momentum range
considered, the \(J/\psi\) meson exhibits \(\rho_{00}^{h}>1/3\),
whereas the \(\phi\) meson has \(\rho_{00}^{h}<1/3\),
corresponding to preferences for longitudinal and transverse
polarization, respectively. For the \(J/\psi\), increasing the
rotation parameter generally enhances the positive deviation from
\(1/3\) at low and intermediate momenta for propagation along
\(z\), but can suppress it at high momenta. For propagation along
\(x\), increasing the rotation parameter has a small effect at low
momenta and slightly reduces the positive deviation from \(1/3\)
at high momenta. For the \(\phi\) meson at the same temperature,
increasing the rotation parameter generally reduces
\(\rho_{00}^{h}\) at intermediate and high momenta, with a more
pronounced effect for propagation along \(z\). These results
reveal a directional dependence of the spin alignment in the
rotating background.

Increasing \(T\) from \(0.15\) to \(0.20~\mathrm{GeV}\) reduces
the positive deviation for the \(J/\psi\). For the \(\phi\) meson
at \(T=0.10~\mathrm{GeV}\), the deviation depends nonmonotonically
on momentum and changes from positive to negative at larger
momenta. At \(T=0.15~\mathrm{GeV}\), it is negative over most of
the momentum range considered.

We further studied the global spin alignment \(\rho_{00}^{\,y}\)
defined along a fixed quantization axis and investigated its
azimuthal-angle and rapidity dependence. At
\(T=0.15~\mathrm{GeV}\) and \(p_T=2~\mathrm{GeV}\), the two mesons
exhibit opposite azimuthal patterns at midrapidity \(Y=0\):
\(\rho_{00}^{\,y}\) reaches a maximum near \(\varphi=\pi/2\) for
the \(J/\psi\) and a minimum for the \(\phi\). At the same
temperature and transverse momentum, increasing \(a\) at \(Y=1\)
decreases \(\rho_{00}^{\,y}\) for the \(J/\psi\) but increases it
for the \(\phi\). At the same temperature, for \(p_T=1\), \(5\),
and \(10~\mathrm{GeV}\), after averaging over the azimuthal angle
with the elliptic-flow weight, the global spin alignment of the
\(J/\psi\) decreases with increasing rapidity within the range
considered, while that of the \(\phi\) increases. The deviations
of the global spin alignment for the \(\phi\) can reach the
\(\mathcal{O}(10^{-2})\) level, comparable in magnitude to those
observed by the STAR Collaboration~\cite{Abdallah_Nature2023}. The
resulting azimuthal-angle and rapidity dependence can be compared
with future measurements at RHIC and the LHC within the kinematic
range considered in this work. In particular, a simultaneous
measurement of the sign of the azimuthal modulation near
\(\varphi=\pi/2\) and of the rapidity ordering for the two mesons
would provide a direct test.

The main result is a systematic comparison of two mesons in one
rotating background with one spectral framework, together with a
map of the momentum, rapidity, and azimuthal
dependence. This comparison provides a reference for future tests.
If future data in the relevant kinematic range show trends
opposite to those found here, the simplest single homogeneous
rotating-medium interpretation would be challenged. Agreement would support the compatibility of the observed trends
with this homogeneous-background description, without establishing
a local-vorticity origin. This reference is not tied to a particular experimental
energy; it is a statement about how the same rotating strongly
coupled medium imprints different spin density matrices on mesons
of different mass and binding energy.

It is important to distinguish the model-dependent numerical
results from the robust qualitative features. The numerical values
of the resonance peak positions and widths, the precise momentum
at which the \(J/\psi\) helicity-frame deviation changes its
rotation-parameter ordering, and the detailed temperature
dependence are specific to the present soft-wall model, dilaton
profiles, and the instantaneous freezeout approximation. By
contrast, the following qualitative features are stable within the
present holographic model class and form the main qualitative
observations of this paper:
\begin{enumerate}
\item[(i)] a single homogeneous rotating strongly coupled medium
can imprint opposite helicity-frame spin-alignment preferences on
\(J/\psi\) and \(\phi\) over most of the momentum range at
\(T=0.15~\mathrm{GeV}\); \item[(ii)] the rotation-parameter effect
is stronger for propagation along the rotation and beam direction
\(z\) than along \(x\); \item[(iii)] the global spin alignment of
the two mesons shows opposite azimuthal patterns at midrapidity
and opposite rapidity trends after azimuthal averaging;
\item[(iv)] the \(\phi\) global deviation can reach the
\(\mathcal{O}(10^{-2})\) level.
\end{enumerate}
These features provide the main contact with experiment in the
present work. If future data in the relevant kinematic range show
the opposite signs or the opposite rapidity ordering, the simplest
single homogeneous rotating-medium interpretation would be
challenged. Agreement would support the compatibility of the observed trends
with this homogeneous-background description, without establishing
a local-vorticity origin.

Finally, we emphasize that the present analysis is based on a
bottom-up soft-wall holographic model. The planar limit of the MP
AdS black hole is equivalent to a uniformly boosted
AdS--Schwarzschild black brane and does not describe the full
spatial vorticity profile of the QGP produced in heavy-ion
collisions. Moreover, at fixed Hawking temperature, varying the
rotation parameter also changes the temperature in the medium rest
frame. The numerical results should therefore be interpreted
within this homogeneous approximation rather than as isolated
effects of local vorticity. We use an instantaneous freezeout
approximation to relate the spectral functions in the medium to
the final-state dilepton observables~\cite{Sheng_PRD2024}; this
approximation does not describe the dynamical details of the
conversion of in-medium excitations into freely propagating
mesons. Future work may consider spatially nonuniform rotating
backgrounds and a more detailed description of the freezeout
process.

\section{Acknowledgments}
This work is supported by the National Natural Science Foundation
of China (NSFC) under grant No.~12375140.

\clearpage

\end{document}